# Human–AI Co-Performance in Fitness Education: A Longitudinal Study with a Pilates Instructor

Qian Huang, Lee Kuan Yew Centre for Innovative Cities, Singapore University of Technology and Design, Singapore. Email: qian_huang@sutd.edu.sg

Poon King Wang, Lee Kuan Yew Centre for Innovative Cities, Singapore University of Technology and Design, Email: poonkingwang@sutd.edu.sg

## Abstract

Artificial intelligence (AI) is poised to transform teaching and coaching practices, yet its optimal role alongside human expertise remains unclear. This study investigates AI–human co-performance in fitness education through a one-year qualitative case study with a Pilates instructor (the "Pilates Instructor"). The researcher engaged as a participant-observer in the instructor's classes (2–3 sessions per week, totaling over 200 hours) and conducted biweekly semi-structured interviews (approximately 60–90 minutes each,30 in total) to explore how generative AI (GPT-4) could be integrated into class planning and instruction. Three research questions guided all observations and interviews: (1) In what situations do humans perform better than AI? (2) In what situations does AI perform better than humans? (3) In what situations do humans and AI perform better together? Findings indicate that GPT-4 excelled at technical and information-intensive tasks – providing rapid movement analysis, diverse exercise suggestions, and polished cueing – whereas the human instructor surpassed AI in embodied demonstration, emotional support, and adaptive creativity. Importantly, the combination of AI and human strengths yielded improved outcomes: co-designing class routines, augmenting feedback with AI's insights and human judgment, and personalizing instruction at scale. These results, illustrated through case vignettes (e.g., exploring how to teach Pilates breathing and diagnosing a student's toe pain), suggest that AI functions best as a complementary tool or "teammate" to the instructor rather than a replacement. The paper concludes with implications for educational technology and sports science: by harnessing AI's capabilities while preserving the indispensable human touch, teaching effectiveness and student experiences in fitness settings can be enhanced.

## Introduction

Generative AI technologies such as OpenAI's GPT-4 have rapidly entered the education landscape, raising questions about how they can support or augment human instructors. Early evidence in academic settings shows that AI tools like ChatGPT can reduce teachers' workload in tasks like lesson planning, allowing educators to generate quizzes, brainstorm activities, and customize materials more efficiently (Baxter,2024). In one trial, teachers who used ChatGPT cut their planning time by about 31% compared to peers who did not use AI. Such productivity gains illustrate what AI can do well – e.g. automate routine content creation – freeing human teachers to focus on higher-level pedagogical work (Kuka & Sabitzer, 2024). At the same time, there is caution that AI outputs must be vetted for quality and that the irreplaceable aspects of teaching (such as mentorship and ethical judgment) remain with humans (Pashaie, Mohammadi, & Golmohammadi, 2024). This balance is especially pertinent in fitness education, a domain that combines cognitive learning with physical, embodied practice.

In fields like sports coaching and physical training, AI is beginning to be used for data analysis and personalized training programs. For instance, AI systems can analyze athletes' performance metrics or even video of movements to provide feedback or suggest improvements. However, coaching research emphasizes that the "human touch" – leadership, motivation, and individualized support – is critical for athlete development. AI might excel at crunching data or suggesting optimizations, but maintaining the human elements of empathy, creativity, and ethical oversight remains a "critical challenge" in integration. Pashaie et al. (2024) describe a *"Synergy Theory"* in sports coaching, wherein AI enhances training outcomes only when combined with human-centered practices, cautioning against over-reliance on AI that could undermine the coach-athlete relationship. This resonates with broader debates on whether AI should be viewed as just a tool or as a collaborative teammate in professional work. Some argue that even advanced AI should *not* be anthropomorphized as a teammate, as it lacks genuine understanding or agency. Others explore the potential of human–AI teaming, suggesting that under the right design conditions, AI can function as a partner supporting human goals (Seeber et al., 2020). In practice, the line between tool and teammate may be blurred – an AI system like GPT-4 can take initiative in generating ideas (teammate-like), yet ultimately it operates under human prompts and validation (tool-like).

This paper seeks to empirically examine these issues in the context of fitness instruction. Specifically, we report on a year-long class observation (2-3 times per week night classes, around 200 hours in total) and rounds of interviews (bi-weekly, around 40hours in total) with a Pilates and yoga instructor, and her exploratory use of GPT-4 as an instructional aid. Pilates is a movement discipline emphasizing form, alignment, and mind–body awareness, where instructors commonly use demonstration, tactile feedback, and real-time cueing. How might a text-based AI system fit into this practice? What aspects of teaching can the AI augment, and what aspects resist automation? By immersively studying The Pilates Instructor's integration of GPT-4 into her classes, we aim to identify what AI can do better, what humans do better, and what both can do better together in fitness education. These research questions are grounded in real-world practice, with qualitative evidence (including case study vignettes) illustrating the evolving dynamics of AI–human collaboration. The goal is to move beyond hype or fear of AI in education, and instead offer a nuanced understanding of complementary strengths – ultimately informing how instructors in sports and educational technology can harness AI as a powerful ally while safeguarding the irreplaceable qualities of human teaching.

**Literature Review**

*AI in Educational Practice*

The advent of large language models (LLMs) like ChatGPT has generated both excitement and concern in education. On one hand, AI offers unprecedented access to knowledge and rapid content generation, which teachers can leverage for lesson planning, creating learning materials, or receiving tutoring suggestions[1]. Studies have found that when used judiciously, generative AI can *"revolutionize classroom dynamics"* by alleviating administrative burdens and providing personalized resources. For example, an autoethnographic study by Kuka and Sabitzer (2024) highlighted AI's pivotal role in reducing a teacher's workload (e.g., drafting curriculum plans), allowing more time for direct student engagement. On the other hand, educators and researchers stress the need for critical oversight: AI-generated content may contain errors or biases, and ethical issues arise regarding transparency (should teachers disclose AI involvement?) and fairness. There is also a psychological dimension – over-reliance on AI or anthropomorphizing it could lead to misplaced trust. Thus, scholarly consensus suggests AI be used as an assistive tool, with educators maintaining *"human-in-the-loop"* control to ensure pedagogical soundness. This perspective aligns with human–computer interaction frameworks like Human-Centered AI, which advocate designing AI systems that augment rather than replace human users.

*AI in Sports and Fitness Coaching*

In parallel, the sports science domain has begun exploring AI integration. Applications range from computer vision systems that evaluate exercise form, to machine learning models that tailor training programs for individual athletes (Mateus et al., 2024). Early results are promising in areas like performance analysis, injury risk prediction, and optimizing workout regimens. For instance, AI can digest large volumes of biomechanical data far faster than a human coach, spotting patterns or inconsistencies that might be missed. However, researchers emphasize that effective coaching is not just about data – it requires mentorship, motivation, and adaptability. Pashaie et al. (2024) note that *"maintaining the human touch in coaching remains critical"* even as we leverage AI's capabilities. Their *"Synergy Theory"* model proposes that the best outcomes occur when AI and traditional coaching methods are combined, highlighting ethical standards and human-centered practices as non-negotiable. Over-reliance on AI, conversely, could erode the coach–athlete bond and lead to issues of trust and compliance. These insights from sports engineering literature suggest that co-performance – AI and humans working in tandem – should be the design goal, rather than one displacing the other. They also underscore challenges such as data privacy, the need for coaches to be literate in AI tools, and the importance of context (e.g., AI might excel with elite athletes' data, but a novice exerciser might need the encouragement only a human can give).

*Conceptualizing AI's Role: Tool, Teammate, or Neither?*

A useful lens for our investigation is the question of how practitioners perceive and use AI: as a tool (an advanced instrument fully under the user's command), as a teammate (a semi-autonomous partner contributing ideas and feedback), or perhaps *neither* in a clear-cut way. In educational technology discourse, there is a spectrum of viewpoints. Traditionalists maintain that *"machines are tools"* and caution against attributing human-like agency to AI. In this view, AI lacks intentionality and should be treated as a sophisticated aid that enhances human labor but does not fundamentally share in the creative or decision-making process. Progressive voices in HCI and AI research explore *human–AI teaming*, where AI systems are designed to collaborate, communicate, and even coordinate with humans in more egalitarian ways (Seeber et al., 2020). Early experiments in domains like programming, writing, and medicine show AI can sometimes take on proactive roles (suggesting solutions, double-checking human decisions, etc.), effectively acting like a "junior partner" on a team. That said, even proponents of AI teaming acknowledge current AIs do not possess true understanding or accountability. Thus, a pragmatic stance emerging in literature is to leverage AI for what it's best at (pattern recognition, information retrieval, speed and scale of output) while relying on humans for what they uniquely offer (judgment, empathy, tacit knowledge). This study adopts that pragmatic stance: rather than debating terminology, we empirically examine the division of labor between AI and a human instructor in a real teaching scenario. The "Tool, Teammate, or Neither" framework serves as an undercurrent to our analysis, helping interpret how The Pilates Instructor at times used GPT-4 as a tool (e.g., a convenient source of ideas), and at other times interacted with it in a more teammate-like fashion (e.g., iteratively refining class plans, almost like brainstorming with a colleague). Ultimately, our findings are structured around specific capabilities (per the three RQs), to concretely show where AI or human had an edge and where a symbiotic partnership emerged.

---

[1] educationendowmentfoundation.org.uk

**Methodology**

*Data Collection Procedures*

Observations: Fieldwork spanned approximately 12 months ( January–December 2024). The researcher joined the Pilates Instructor's classes regularly, about 2–3 times per week, as a participant-observer. In practice, this meant the researcher was often an active participant in the exercise class (performing the exercises alongside other students) while also taking mental or discreet written notes on notable events. Immediately after each class, the researcher would finalize detailed field notes describing what occurred. These notes focused on the instructor's teaching behaviors, student reactions, and any situations that stood out as challenging, innovative, or relevant to the research questions. In particular, the observer paid attention to moments that might represent potential opportunities or pitfalls for AI assistance. For example, if during a class a student asked an unusual question about anatomy that the instructor had to pause and think about, this would be noted as a scenario where an AI might provide quick information. Similarly, if the instructor spontaneously adapted an exercise to accommodate a student's injury or used humor to motivate the class, such instances were recorded as examples of human strengths that AI might lack. Because the instructor was not actively using GPT-4 in class, these observations helped identify hypothetical points of insertion for AI – junctures where we could later ask, "Could AI have helped here? How or why not?"

Interviews: Roughly every two weeks, the researcher conducted a semi-structured interview or extended debrief with the Pilates Instructor (usually after a class or at a convenient break time). These interviews lasted about 60–90 minutes each and ultimately totaled approximately 30 sessions over the year. The interviews were conversational in nature, often taking place in a quiet corner of the studio or at a nearby café right after class. An interview guide ensured that certain key questions related to the research questions were covered. We repeatedly asked the instructor to reflect on RQ1, RQ2, and RQ3 in the context of recent experiences. For instance, she was asked: "Looking back at this week's classes, were there moments you felt you handled something much better than any AI could?" (addressing human strengths), and conversely, "Were there any moments you wish you had some AI help, or think an AI might have done a better job at something?" (addressing AI strengths). We also discussed hypothetical scenarios: "If GPT-4 were available at that moment, how do you imagine using it?" and "Do you think having an AI suggest something in that situation would have been useful or not?" Additionally, the instructor was prompted to describe any synergistic ideas – for example, could she envision a way that both she and an AI might work in tandem during an aspect of teaching (planning, demonstration, feedback, etc.) to improve outcomes? These questions were kept broad to allow the instructor to bring up her own insights and experiences.

During many of these interview sessions, the researcher and instructor also engaged in hands-on exploration of GPT-4's capabilities. The researcher had a laptop or tablet with access to GPT-4 (via a ChatGPT interface that could handle text and, in some cases, images for analysis). In a collaborative spirit, they would sometimes test the AI in real time. For example, after a class in which a student had a particular problem (say, difficulty with balance in a certain pose), the instructor might wonder if AI could offer any suggestions. The researcher would then input a description of that scenario into GPT-4 (and occasionally even share a photo or video snippet from the class, with participants' consent, to see what the AI would say). The instructor would review the AI's output and discuss her impressions: Did the suggestion seem useful? Was it something she hadn't thought of? Would she consider using that advice in a future class? This interactive experimentation allowed the instructor to experience AI assistance in a low-stakes setting, and her reactions and reflections were recorded. For instance, in one interview, the instructor asked GPT-4, "How can I help students understand proper Pilates breathing while standing?" (a scenario we detail in the findings). In another, she described a student's knee pain issue and we examined GPT-4's recommendations for modifications. These dialogues with the AI were captured either by saving the chat transcripts or by taking notes on the AI's answers and the instructor's responses.

*Data Analysis*

All qualitative data (field notes, interview transcripts, message logs, GPT logs) were imported into a qualitative analysis software for coding. The analysis was thematic, guided primarily by the three research questions. We utilized both a priori codes (e.g., "AI strength" vs "Human strength" vs "Synergy") and emergent codes that arose from the data (e.g., a code for "musicality" emerged when multiple instances showed The Pilates Instructor's use of music in ways AI couldn't match; a code for "new idea from AI" captured moments when GPT introduced a novel insight, etc.). The researcher and a peer qualitative analyst independently coded a subset of data to refine the codebook, improving trustworthiness through intercoder agreement on key themes. Triangulation was achieved by comparing insights from different data sources – for example, if a GPT log showed a particular suggestion and interview data indicated whether The Pilates Instructor tried that suggestion in class, we looked at student reactions in observations to form a holistic view.

**Findings**

We organize the findings around the three research questions, each corresponding to a set of themes that emerged from the data. To illustrate these findings, we interweave **case examples** drawn from the interview narratives – in particular, two vignettes: one involving The Pilates Instructor using GPT-4 to diagnose a student's movement problem ("tuck the toe"), and another involving GPT-4 assisting in teaching Pilates breathing. These cases serve as concrete examples of AI's capabilities, human expertise, and their interplay.

**Case Study 1:** *Co-Teaching Pilates Breathing with AI Assistance*

One of the earliest and illuminating instances of human-AI collaboration we observed was during The Pilates Instructor's effort to improve how she teaches Pilates breathing, a foundational skill in Pilates that involves coordinated ribcage expansion and deep diaphragmatic breathing. Pilates breathing technique can be challenging for beginners to grasp because it involves subtle internal actions (engaging intercostal muscles, expanding the ribs laterally) that are not easily seen. The Pilates Instructor wanted to refine her approach to teaching this in her group classes. She decided to consult GPT-4 for ideas and best practices, effectively bringing an AI "brainstorming partner" into her lesson planning.

In a planning session, The Pilates Instructor typed a prompt into GPT-4 asking how to best teach Pilates breathing to students. She described her class context and asked for suggestions, including any visualizations or exercises that could help students understand the breath. GPT-4 responded by outlining what it called a "classic method" for introducing Pilates breathing, recommending that new learners first practice the breathing lying down on a mat, where they could better feel their back expanding against the floor. The AI's response even included a brief description of an illustration: it suggested imagining or showing an image of a person lying supine with hands on their ribs, indicating inhalation/exhalation movement. Essentially, GPT-4 drew from standard Pilates pedagogy in its training data – which often begins breathwork in a supine position – and presented that to The Pilates Instructor. It also offered a couple of YouTube links to videos demonstrating Pilates breathing fundamentals, providing The Pilates Instructor with external resources at the click of a button. In this phase, GPT-4 was acting as a *knowledge tool*, retrieving established techniques and resources. The Pilates Instructor acknowledged that the AI's recap of the traditional approach was useful ("It's basically confirming the textbook approach," she noted), and she watched one of the suggested video links, which she found to be a good refresher on the mechanics of ribcage expansion.

However, The Pilates Instructor's own teaching experience led her to adapt these suggestions. In a follow-up exchange with GPT-4, she mentioned that her classes are group fitness settings where students typically stand or sit, not lie down, for warm-ups. She queried if there were ways to teach the same breathing concept in a standing position, which would be more practical in her classes. GPT-4 duly came up with ideas: it suggested having students stand back-to-back to feel each other's ribcage expansion, or using resistance bands around the ribcage as a tactile cue while standing. These ideas were creative, but The Pilates Instructor was unsure how well they would work with a large group. Ultimately, she decided on her own pedagogical modification: she would teach the breathing in a standing position by demonstrating on herself and then using peer observation. Specifically, in class she had students stand in a circle, placed one student's hands on her (The Pilates Instructor's) ribcage so they could feel her ribs expand as she inhaled, and then encouraged them to pair up and feel each other's ribcage movement. This was something The Pilates Instructor came up with – inspired partly by the principle of tactile feedback that the AI's suggestions reinforced, but implemented in her own style. During this part of the class, The Pilates Instructor also physically went around gently placing her hands on students' lower ribs and back to guide them ("feel my hands expand as you breathe in") – a very intimate, human form of instruction that no AI could directly provide.

> *A fieldnote excerpt from that session highlights the synergy and limits of the AI's contribution:*
>
> *Fieldnote (Class observation, Apr 5):* The Pilates Instructor brings up the concept of lateral breathing. She says to the class, "We normally teach this lying down, but I want you to be able to do it standing since we'll use it in moves today." She demonstrates an inhale, exaggerating the expansion of her ribcage. She invites one volunteer to place hands on her sides to feel it – a student named M does so and nods in surprise when The Pilates Instructor inhales deeply. The class laughs softly. The Pilates Instructor then has them pair up: "One of you do a big Pilates breath, the other put hands on their ribs to feel it." … *Afterwards*, The Pilates Instructor addresses the class: "Notice how much your ribs can actually move – that's what I want you to use when we do the arm arcs next. That big 3D breath." She circulates, giving a few gentle taps on people's back or sides to encourage more expansion. I note that this physical/tactile element really engages everyone; people seem curious and more confident in trying the breath now. *(Later, in debrief, The Pilates Instructor mentioned that the idea to do it standing came from her past trial-and-error; "Honestly, lying down they get it, but then they stand up and forget. So I skipped straight to standing this time," she said.)*

In discussing this case afterwards, The Pilates Instructor reflected on how GPT-4 was helpful and where it fell short. The AI gave her a starting structure (the supine breathing exercise, which she actually decided not to use in class but it reminded her of fundamentals) and some creative tweaks (back-to-back breathing) that she considered but found impractical. *"GPT-4 basically threw a bunch of ideas at me. They were solid ideas, just not perfectly suited to my situation,"* she said. The final approach she used was something she *customized*, blending AI ideas with her own experience.

**Task 2: Pilate breathing**

**Step 1:** The instructor uses ChatGPT to explore how to better teach students Pilates breathing.
**Step 2:** GPT provides the instructor with a classic method, suggesting learners lie down on the mat and practice breathing techniques (with illustrations included). (AI doing better)
**Step 3:** However, based on the instructor's teaching experience, demonstrating breathing while standing is more effective because, in a group fitness class, learners can observe more clearly. (Human doing better)

(Step 3 and 4: humanXAI doing better)
**Step 4:** The instructor encourages learners to touch her body to feel how the ribcage moves during breathing and also touches the learners' bodies to help them correct their movements. (Human doing better, AI cannot do) (human-human Interaction)

In this case:

AI better than Human:
1) Provide practice classic/traditional routine based on the existing data
2) Provide video links by searching internet

Human better than AI:
1) Teaching experience (domain knowledge): found that letting group learners stand up is better
2) Students still prefer human instructors' communication (human sense/touch)
3) Some students got frustrated/doubtful while learning, human instructor can provide emotional support.

AI and Human works better as ONE:
1) Can provide classic/tradtional way and instructors adapt it to actual situation
2) Can provide basic and instructor is more creative
3) Can provide video links/resources for both instructors and students for their self-directed learning

*Fieldnotes 1: The task of "Pliates breathing"*

**By GPT**

> Can you teach me how to breath in pilates? Pilate breathing

Absolutely! Pilates breathing is quite unique and is known as **lateral breathing** or **ribcage breathing**. Unlike yoga, where breathing can be more diaphragmatic (belly breathing), Pilates breathing focuses on expanding the sides and back of the ribcage while keeping the core engaged. Here's how to practice it:

**Step-by-Step Guide to Pilates Breathing**

1. **Start with Alignment**
   - Sit or lie down in a comfortable position, with your spine long and neutral. If lying down, bend your knees and keep your feet flat on the floor, hip-width apart.
   - Place one hand on your ribcage and the other on your belly to help you feel the breath.

2. **Inhale through the Nose**
   - Take a deep inhale through your nose, but instead of letting the breath fill your belly, try to expand the sides and back of your ribcage.
   - Imagine your ribs are like an accordion, expanding outward and upward with each inhale.
   - Focus on filling the lungs, but avoid lifting your shoulders or arching your back.

3. **Exhale through the Mouth**
   - Slowly exhale through your mouth as if blowing out through a straw. Feel your ribs close and knit together as you breathe out.
   - Engage your core (imagine gently pulling your navel toward your spine) as you exhale, maintaining the ribcage stability while allowing your belly to stay slightly pulled in.
   - This engagement provides stability and helps support your lower back.

**By Human Instructor**

> Actually it is the same as what I'm teaching. I used to start the class laying down on the mat with breathing techniques. But I find students cannot see how my rib cage moves in the lateral way.        4:46 PM

> **You**
> From a student's perspective, I feel that the teacher's demonstrations, and having students touch your bod…

> Yes        4:47 PM

> AI teaching is classic… Is what I learned before        4:47 PM

> AI Very details but students may not understand. Because I was very blurred when I just started to attend the pilates class in amore as a member. I don't understand what a plevic tite or the name of muscle etc…
>        Edited 4:49 PM

*Fieldnotes 2: Comparison of GPT feedback and Human Instructor's feedback*

From an analytic standpoint, this Pilates breathing case illustrates a clear division of labor and complementary strengths: GPT-4 excelled at providing canonical knowledge and quick

resources – it acted like a tireless research assistant, pulling a standard solution and augmenting it with videos and even suggested imagery. For an instructor like The Pilates Instructor, who already understands the pedagogy, this saved time (she didn't have to dig through manuals or YouTube herself; GPT-4 surfaced relevant content in seconds). Indeed, GPT's ability to rapidly fetch such resources was noted repeatedly by instructors as a major advantage, with one noting how *"GPT delivers relevant YouTube tutorials or anatomy info on demand, which is great when I need to show or learn something quickly"*. In this case, GPT-4's pointers to *classic routines and references* effectively gave The Pilates Instructor a menu of options.

On the other hand, the human instructor's strengths were evident in how the plan was implemented. The Pilates Instructor's *teaching experience and intuition* told her that a standing approach would work better for her class format (an insight GPT-4 did not have, since it lacked the situational awareness of her specific class setting). This aligns with a broader pattern we observed: human instructors excel at context-specific adaptation – knowing their particular students, environment, and what has or hasn't worked in practice before.

> *As The Pilates Instructor noted, "I've tried teaching breathing lying down, but in my classes people then struggled to translate it when standing. So I trust what I've seen."*
>
> That kind of experiential knowledge – gained from trial-and-error and observing student reactions – is something AI lacks. Moreover, the embodied demonstration and tactile guidance The Pilates Instructor provided (inviting students to touch her ribs, and she theirs) is inherently beyond AI's current capabilities. It requires a human presence, trust, and physical interaction. This aspect was crucial for student learning in this case: indeed, students reported after class that the moment of feeling the instructor's ribcage move was an "aha" moment that finally made them understand how to do it. The importance of such embodied, hands-on teaching methods came up frequently. Instructors pointed out that physical feedback – like a teacher placing a hand to guide a limb or, as in this case, to let a student feel a bodily movement – has no real AI *substitute* in live classes. As one instructor put it, *"Correcting a learner's posture often means physically guiding them. You can't have a chatbot poke someone's ribcage"* – highlighting the irreplaceable nature of human touch in movement education.

Crucially, the combination of AI and human input led to a better outcome than either alone in this scenario. GPT-4 contributed a solid foundation (ensuring no known technique was overlooked and enriching the plan with multimedia resources), and The Pilates Instructor contributed the creative adaptation and delivery. Neither acting alone would likely have achieved as effective a result: without AI, The Pilates Instructor might have still taught effectively (she had done so before), but she might not have had the time to gather those extra video resources or consider alternate tactics like back-to-back practice. Without The Pilates Instructor's human insight, an AI-generated plan might have stayed at the generic level (perhaps instructing learners to lie down, which in her context would not have been feasible or as engaging). This synergy echoes a pattern summarized by The Pilates Instructor: *"GPT can provide the classic/traditional way, and then I as the instructor adapt it to the actual situation"*. Indeed, instructors frequently mentioned that AI plus human together works best when the AI provides a baseline or ideas and the human refines them with creativity and contextual understanding. Table 1 (later in this section) will further detail such divisions of labor.

It's worth noting that in this case study, GPT-4 functioned largely as a tool from The Pilates Instructor's perspective – a source of information and suggestions she controlled and picked from. There was an element of back-and-forth (she asked follow-up questions), which edges toward a more interactive, teammate-like engagement, but ultimately The Pilates Instructor made the decisions; GPT-4 did not actively take initiative beyond answering prompts. In her words, *"It's like a brainstorming buddy, but I'm definitely the one driving the process"*. This characterizes many instances we saw: instructors appreciated AI as a brainstorming partner up to a point, yet they maintained a clear sense of *human leadership* in the planning. We will revisit this notion of AI-as-assistant vs AI-as-partner in the discussion of the Tool/Teammate framework.

**Case Study 2:** *Diagnosing a Movement Problem – "Tucking the Toe"*

A second case highlights how AI and an instructor can jointly tackle a technical problem in movement execution, and also the challenges when AI advice clashes with human experience. This scenario unfolded during a Pilates reformer training session where Researcher experienced pain in her big toe while attempting an exercise that involved tucking the toes (flexing the toes under, as in a plank or kneeling position). The case provides a window into how AI's analytical precision can complement the human trial-and-error approach – and how instructors vet AI's unconventional suggestions.

During a practice session, Researcher repeatedly felt a pain at the base of her big toe whenever she did a certain lunge movement with toes tucked. Puzzled, she asked The Pilates Instructor, *"Why does my toe hurt so much in this position?"* The Pilates Instructor's initial hypothesis was that it could be due to limited flexibility in Researcher's feet or ankles. Acting on this hunch (a common one – tight toe extensors or arches can cause discomfort in such poses), The Pilates Instructor had Researcher do some related stretches and modifications: pointing and flexing the foot, using a smaller range of motion, etc. They spent several minutes experimenting, but Researcher still felt the pain when returning to the original position. The Pilates Instructor then stepped back and observed Researcher's foot closely as she attempted the toe-tuck. At this point, the researcher suggested using GPT-4's diagnostic capability as an aid. They snapped a quick photo of Researcher's feet in the problematic position and fed it to GPT-4 with a question describing the issue ("This student feels pain in the big toe when doing a toe-tuck lunge – any idea why?").

GPT-4 analyzed the image and the question. Impressively, it identified a specific detail: *the ball of Researcher's foot (the pad behind the toes) was not making full contact with the floor.* The AI hypothesized that because of this, when Researcher's weight was on the toes, the pressure might be concentrated improperly on the big toe joint, causing pain. Essentially, GPT-4 pinpointed a form issue that neither The Pilates Instructor nor Researcher had noticed yet – that Researcher's foot alignment was slightly off, likely due to anatomy or flexibility, and her weight distribution was suboptimal. This turned out to be exactly the cause of the pain. When The Pilates Instructor looked again with this insight, she realized that indeed Researcher's toes were bending but her foot was not "planting" down fully; a portion of her forefoot was lifted.

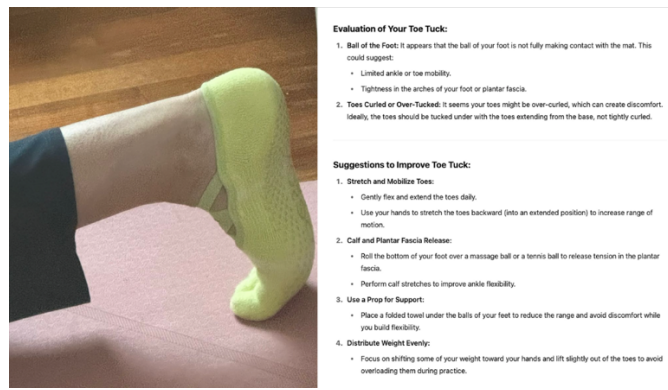

*Fieldnote 3: GPT's feedback for the task of "Tucking the Toe"*

We later learned that GPT-4's identification corresponded to a known phenomenon: if the metatarsophalangeal joints (ball of foot) don't bear weight, too much stress goes to the distal phalanx of the big toe. This kind of diagnostic reasoning is a strength of AI – scanning visual input for alignment clues and matching to its knowledge of biomechanics. As one instructor noted, *"AI can analyze posture from photos and catch things we might miss"*, giving the example of detecting that a foot was not properly grounded and linking it to toe pain. In our study, this was a clear instance of AI outperforming the human in immediate technical analysis: The Pilates Instructor admitted she might have eventually caught the issue, but the AI's quick answer saved time. Indeed, GPT-4's precision in movement diagnostics was noted in multiple trials; it could, for example, identify subtle asymmetries or incorrect angles in images of yoga poses and suggest corrections, functioning almost like a virtual movement coach. This aligns with emerging uses of AI in sports science for posture analysis

Armed with the AI's diagnosis, The Pilates Instructor and Researcher addressed the issue. The Pilates Instructor instructed Researcher on how to adjust her foot placement – basically to consciously press the ball of the foot down. Researcher tried this and felt an improvement: the toe pain reduced significantly. In this sense, the human and AI worked *together* to solve the problem: AI identified the cause, and the human implemented the solution with the student.

The interaction didn't end there, however. GPT-4 had also offered a specific *solution* in its response: it suggested that placing a rolled towel under the toes during the movement could alleviate the pressure on the big toe. This was an intriguing recommendation that neither The Pilates Instructor nor Researcher had heard of. The Pilates Instructor was initially skeptical – in her 18 years of teaching Pilates and related disciplines, she hadn't encountered the "towel under toes" trick. It seemed a bit strange, but not impossible. They decided to experiment: Researcher placed a small towel roll under her toes as GPT-4 suggested and performed the lunge. Unfortunately, this didn't help; in fact, Researcher reported it felt even more awkward, as the towel reduced her stability and she couldn't feel a solid base ("I can't feel the support" she said, meaning her foot had less feedback from the firm ground). The Pilates Instructor chuckled and remarked that the towel idea was "creative, but probably not practical here." She then offered an alternative fix from her own repertoire: she had Researcher perform the move with a slight external rotation of the hip (turning the leg out a bit) and using a foam cushion under the *knee* instead, to change the angle. This was a modification The Pilates Instructor knew could sometimes relieve toe/ankle strain. That approach ultimately resolved Researcher's issue – the pain went away when she did the movement with that modification.

This portion of the case reveals an important dynamic: AI might propose solutions that are theoretically sound but lack the practical wisdom that experienced instructors develop. GPT-4's towel suggestion was not wrong per se – it likely surmised that cushioning could help – but it failed to account for how that would feel or what secondary problems it might create (loss of stability or proper feedback). The Pilates Instructor's reaction ("never heard of this way… [it] was not effective, I used another way") underscores that the instructor's expertise was crucial to evaluate AI's idea. She effectively vetted the AI's suggestion against her domain knowledge, a pattern we saw frequently. Instructors did not take AI output as gospel; rather, they tested and filtered it. In interviews, they expressed a healthy skepticism: *"Sometimes GPT suggests an odd fix. If it sounds off to me, I trust my gut or double-check with my knowledge"* – a quote reflecting The Pilates Instructor's mindset in this scenario. This highlights how human instructors remain the ultimate decision-makers, ensuring safety and suitability of any advice.

Despite the towel idea flop, Researcher and The Pilates Instructor both found value in GPT-4's assistance. GPT's other contribution in this case was providing additional learning resources after addressing the immediate issue. It offered a list of exercises and stretches to improve toe and foot flexibility, complete with YouTube video links (for example, a link to a foot mobility exercise from a physical therapy channel). Researcher later watched these videos and said she found them "quite useful" for her own knowledge. This points to another role of AI: not just solving the problem at hand, but enhancing the instructor's (or student's) ongoing learning by curating relevant content. In this sense, GPT-4 acted like a personalized tutor for Researcher, guiding her to work on her foot flexibility which could prevent future problems. It exemplifies how AI can extend an instructor's capability by supplying on-demand educational material for further practice. One instructor noted that GPT-4 effectively became a "virtual library" for them, pointing them to well-regarded sources (e.g. GPT often recommended the *Muscle and Motion* YouTube channel for anatomy and movement demonstrations, which our instructors came to appreciate.

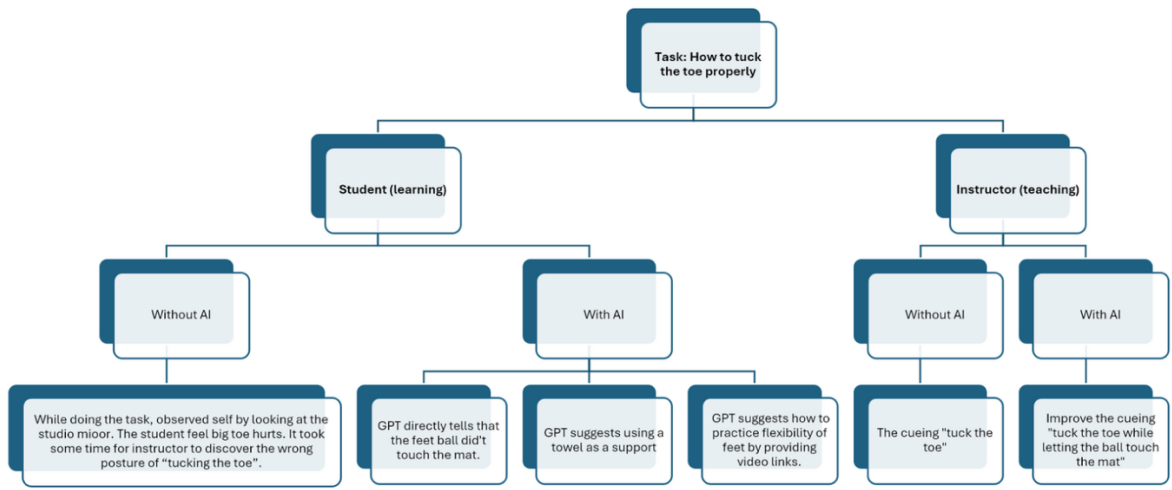

*Figure 1: The human-AI collaboration for the task of "Tucking the Toe"*

To generalize insights from the cases above and numerous other instances in our data, **Table 1** presents a summary of what humans do better, what AI does better, and what they achieve best together in the context of fitness education. This synthesis draws on recurrent themes from interviews and observations, with supporting examples (some already described, others from additional field cases):

| Aspect of Teaching | Human Instructor Strengths | AI (GPT-4) Strengths | Combined (Human + AI) Synergy |
|---|---|---|---|
| **Movement & Posture Diagnosis** | - Observing nuanced body language and sensing kinesthetic feedback in real-time (e.g. feeling muscle tension via touch). <br> - Inferring causes of issues through experience ("I've seen this kind of imbalance before"). | - Analyzing images or descriptions to pinpoint technical issues with precision (e.g. identifying a misaligned foot causing toe pain) . <br> - Cross-referencing vast knowledge of anatomy and common errors (never "forgets" rare conditions). | - Human + AI can jointly assess movement: AI provides an initial diagnosis or possibilities, and the instructor confirms via hands-on checking and questions the student. *Example:* AI flags a knee valgus in a squat from a photo; the instructor then physically checks the student's knee tracking and addresses it. |
| **Designing Exercise Routines** | - Creative sequencing with *musicality* and flow tailored to the class energy.<br>- Adapting on the fly to participant needs (substituting exercises if someone has an injury, adjusting intensity in real-time).<br>- Drawing from personal style or local culture (e.g. incorporating dance elements unique to the instructor's background). | - Generating a variety of routine options quickly (dozens of exercise combinations for a given goal).<br>- Ensuring all muscle groups or training components are covered (doesn't inadvertently omit something).<br>- Providing structured progressions (e.g. from beginner to advanced variations) based on learned guidelines. | - **Co-creation of routines**: AI suggests a "starter" routine or menu of moves; instructor curates and rearranges them to create a coherent, engaging class sequence.<br>- AI ensures no critical element is missing (serving as a checklist), while instructor ensures the routine has rhythm and suits the actual group.<br>*Example:* For a strength yoga class, GPT-4 proposes 15 poses; the teacher selects 10, changes the order for better flow, and adds music that matches each segment. |
| **Instructional Cues & Explanations** | - Using intuitive analogies and metaphor drawn from personal experience to convey movement quality ("Imagine you're balancing a book on your head") – often tuned to the audience's reactions.<br>- Reading the room and providing motivational energy, empathy, or humor to keep learners engaged.<br>- Multimodal communication: demonstrative gestures, tone of voice, and touch. | - Providing clear, concise technical descriptions (e.g. step-by-step breakdown of an exercise) and anatomical details on demand.<br>- Suggesting alternative phrasing for cues, including ones the instructor hadn't thought of, to clarify instructions.<br>- Instantly answering factual questions (e.g. "Which muscle does this exercise target?") with accuracy. | - **Refined cueing**: Instructor and AI iterate on how to phrase cues – the instructor describes what they want to convey, AI offers polished wording or additional analogies, and the instructor delivers the final cue with personal flair.<br>- AI can supply scientific context or novel metaphors which the instructor then translates into a relatable form for students.<br>*Example:* An instructor asks GPT-4 for a metaphor to teach core stability; AI suggests "tighten your core as if bracing for a punch." The instructor tries this cue in class, sees improved student understanding, and adds an encouraging tone while saying it. |
| **Emotional Support & Motivation** | - Sensing student frustration or confusion and responding with encouragement or adjusted instruction on the spot.<br>- Building rapport and trust, which increases student confidence and motivation ("I know you can do this, I've seen your progress").<br>- Inspiring through personality – e.g. an instructor's enthusiasm or empathy can galvanize a group. | - Available 24/7 to provide answers or guidance (consistency and patience, albeit without genuine emotion).<br>- Can offer scripted motivational lines or positive affirmations if prompted (though these may feel generic).<br>- Not subject to fatigue – will tirelessly re-explain or give feedback as often as asked, which can help self-directed learners. | - **Augmented coaching**: AI can equip instructors with personalized pep talks or psychological strategies drawn from sports psychology research, which the instructor can then deliver in a heartfelt way.<br>- Outside class, a combination of human and AI support might be offered: e.g. AI chatbot for routine check-ins and the human for deep motivation.<br>*Example:* An instructor uses GPT to generate a few encouraging phrases tailored to common yoga struggles (like balance wobbles), mixes those with her own words, and uses them during class when she notices students struggling – providing timely uplift. |
| **Personalization & Adaptation** | - Knowing individual students' bodies, histories, and personalities, and tailoring guidance accordingly ("I recall you have a shoulder issue, let's | - Absorbing large amounts of personal data (if provided) and remembering it consistently – e.g. could keep track of each | - **Optimized personalization**: The instructor can consult AI for ideas on how to modify exercises for unique needs, then apply the appropriate one with their judgment file-k23gyh9tpfauk5vvbgqrvv |

| | | |
|---|---|---|
| modify this move for you"). <br>- Picking up subtle cues of discomfort or fatigue in a student and adjusting the class plan immediately. <br>- Innovating on the spot if equipment is missing or environment changes (e.g. too crowded, so altering exercises). | student's progress, limitations, and suggest specific modifications. <br>- Offering a library of variations for any given exercise, which the instructor can match to student needs. <br>- Simulating different scenarios ("If student is X tall or has Y condition, what modification to use") and giving quick answers. | . <br>- Over time, AI could learn an instructor's student profiles and pre-generate personalized options, which the instructor fine-tunes. <br>*Example:* For a very tall student struggling with a reformer exercise, the instructor asks GPT for adjustments; AI suggests a foot bar change and a strap length modification. The instructor tries these and finds the student now performs comfortably. |

**Table 1: Comparative strengths of human instructors and AI (GPT-4) in various aspects of fitness teaching**

## Discussion

This one year class observation and rounds of semi-structured interview provides a grounded look at the interplay between a human instructor and an AI (GPT-4) in the context of fitness education. The findings illustrate a nuanced reality: AI and humans have distinct strengths and limitations, and the most promising approach is to integrate them in complementary ways. In this section, we reflect on the broader implications of these findings, connect them to existing frameworks, discuss practical and ethical considerations, and acknowledge the study's limitations and contributions.

Our study sheds light on the oft-posed question of whether AI is merely a tool or can be considered a teammate in professional practice. In The Pilates Instructor's case, GPT-4 was primarily a tool – a resource she used to enhance her work – but at times it exhibited *teammate-like* behavior in that it contributed ideas and feedback in an interactive loop. Crucially, GPT-4 never acted autonomously in the classroom; it did not initiate actions or deviate from what it was asked. The Pilates Instructor was the one steering every usage of AI. This supports the view that current AI systems, lacking agency and understanding, are not teammates in the full human sense (they do not share accountability or situational awareness). However, from the user's perspective, the experience can feel collaborative. The Pilates Instructor described planning sessions with GPT as if "we" (she and the AI) brainstormed together. This indicates that AI can occupy a middle ground: more interactive than a static tool (because it can respond and suggest in a quasi-conversational manner), yet not a true partner. The practical takeaway aligns with a Human-Centered AI philosophy: treat the AI as a powerful assistant. In our findings, whenever AI was used in a way that approached "teammate," it was within a clearly human-supervised process (e.g., co-creating plans with The Pilates Instructor in charge). We did not encounter a scenario where delegating completely to AI was desirable – for example, The Pilates Instructor never considered letting GPT-4 answer students' questions directly or run any part of the class without her mediation. This reinforces that at current capability levels, AI is best conceptualized as a sophisticated tool that can emulate some supportive teammate functions but ultimately requires human direction. Importantly, keeping this distinction in mind can guard against over-reliance or misplaced trust. If one started treating the AI as an equal peer, one might assume it can handle tasks it actually cannot (like making judgment calls in class). Our study echoes the caution in literature about not over-anthropomorphizing AI assistants in education. At the same time, we saw that collaborating with the AI can indeed "free up time, accelerate productivity" in line with optimistic views of AI teammates in the workplace.

In sum, we advocate a balanced framing: AI as a *"co-pilot"* – it can help navigate and suggest, but the human pilot is ultimately flying the plane. This framing encourages leveraging AI's strengths while maintaining human responsibility and oversight.

Designing AI for Fitness Education:

From a design perspective, our study highlights features that would make AI tools more effective in domains like fitness. One clear need is for multi-modal AI – GPT-4 provided text, but much of fitness instruction is visual/kinesthetic. Integrating GPT's language capabilities with vision (ability to actually *see* a student's movement via camera) would be a game changer. We already did a crude version of this by feeding GPT photos and descriptions, but an AI that could directly analyze live video and give feedback (perhaps in a coach's earpiece or on a screen) would enhance the synergy. Some emerging fitness apps attempt this with pose estimation, but combining it with GPT's explanatory power could provide both detection of issues and human-friendly explanations. However, designers should heed that the instructor should remain in control of how such feedback is relayed to students. The AI might, for instance, flag to the instructor privately, "Student in station 3 has knees caving inward on squats," and the instructor chooses how to intervene. This keeps the human as the intermediary who contextualizes the feedback. Another design insight is the value of a personalized AI profile for instructors. If an AI knows the instructor's typical routines, preferred style, and common student demographics, it can tailor its suggestions better (akin to a human assistant who knows your preferences). In The Pilates Instructor's case, a system that automatically adjusted its outputs to be in line with her "fun and flowy" style (perhaps by learning from her past class plans or even listening to her class music) would have been welcomed. Developers could allow instructors to give feedback to the AI (thumbs up/down on suggestions, or edits that the AI learns from) to iteratively personalize it. This moves towards the idea of an AI *teammate* that adapts to the human – within limits – which our findings suggest is fruitful. We also noted how auditory input (GPT reading aloud) helped engagement for The Pilates Instructor.This hints that AI tools might incorporate text-to-speech to present information to instructors who are on the move or prefer listening (imagine an AI briefing the instructor via an earpiece before class on the plan and key points, almost like a co-instructor prepping). Additionally, since The Pilates Instructor liked to draw sketches of routines and anatomy, a tool that can take a photo of a hand-drawn sketch and *understand* it to some extent (maybe pulling up related exercises or checking against known anatomy) would integrate nicely into her workflow. We actually tried a bit of this: The Pilates Instructor drew a sequence flowchart and we showed it to GPT (via the researcher describing it) to get video suggestions. A more direct image recognition could streamline that. In short, AI design for fitness education should focus on integration – integrating modalities (text, vision, audio), integrating with instructor's personal style, and integrating seamlessly into the instructional environment (perhaps via mobile or wearable interfaces). Ultimately, the design goal is an AI that enhances the human instructor's capabilities without getting in the way. The study provided a user's-eye view of what those capabilities should be (fast analysis, idea generation, feedback support) and what pitfalls to avoid (not undermining the instructor's relationship with students, not suggesting unsafe or very out-of-context things without checks). As AI developers and education technologists collaborate, taking into account these on-the-ground insights will help create systems that truly function as helpful partners in teaching.

Ethical and Social Considerations:

The incorporation of AI into fitness instruction also raises a few broader issues. Privacy is one: if instructors use AI that involves student data (like images or performance metrics), ensuring consent and secure handling of that data is critical. In our case, photos were used with permission in a limited setting. Scale that up to a gym setting with cameras, and one must be transparent with clients about what data is being collected and for what purpose. Bias and accuracy of AI content is another concern. We were fortunate that GPT-4's suggestions in this domain were mostly sound, likely because the body of Pilates/yoga knowledge is well-established. But if AI gave a subtly wrong cue (e.g., an unsafe alignment), a less vigilant instructor might pass it on. That's why we stress human verification. There's also the issue of de-skilling: if an instructor leans on AI too much for answers, do they fail to develop their own expertise? We saw for The Pilates Instructor it actually up-skilled her in many ways, but that was because she was intentional about learning from the AI, not just using it as a crutch. In less controlled scenarios, one could imagine an instructor becoming dependent and not growing certain problem-solving muscles. This suggests a need for monitoring and perhaps even certification on proper AI use – somewhat akin to requiring drivers to learn how to use advanced car features safely. Student perceptions form another angle. If students know an instructor is using AI, does it change their perception of the instructor's competence or the class's authenticity? In our observations, because The Pilates Instructor remained clearly the leader and used AI subtly, students reacted positively or with neutral curiosity. But if, say, an instructor read instructions off a screen verbatim from AI, students might feel shortchanged ("I could have gotten that from an app at home"). The value students find is in the instructor's personal touch, as our findings reinforced. So it's ethically important that AI doesn't reduce the personal engagement students get, otherwise the human instructor's role might indeed diminish in perceived value. Inclusivity is another factor: AI could actually help make classes more inclusive by providing modifications for different needs (as we saw), but if misused, it might standardize things too much. Therefore, guidelines on inclusive prompt usage could be beneficial (e.g., always ask the AI for options suitable for different levels).

**Conclusion**

In summary, this one-year qualitative study reveals a multifaceted picture of AI and human co-performance in fitness education. We posed three questions at the outset: What can AI do better? What can humans do better? And what can they do better together? Our findings, grounded in the daily practice of a The Pilates Instructor adopting GPT-4, can be distilled as follows:

AI can do certain things better. It excels at information processing and generation. In our case, AI quickly diagnosed movement issues, provided a wealth of exercise ideas and technical explanations, and delivered well-crafted cues and learning resources on demand. These capabilities often outpaced what the human could do in the same time, highlighting AI's strength in speed, consistency, and breadth of knowledge. AI essentially acted as a super-charged assistant, handling analysis and content creation tasks that allowed the instructor to operate with greater knowledge and efficiency.

Humans inherently do other things better. The instructor's performance was superior in all matters requiring embodiment, empathy, and nuanced judgment. She provided hands-on corrections, emotional support, and motivational energy that the AI could not mimic. Her creative decision-making and adaptations in the moment ensured that teaching was personalized, safe, and responsive in ways beyond the AI's capability. Human intuition and experience guided the class's tone and adjustments, underscoring that the art of teaching – especially movement teaching – remains a profoundly human skill.

Together, AI and human did even better.The real promise lay in combining their strengths. By working together, The Pilates Instructor and GPT-4 produced better outcomes than either alone – classes were more innovative and tailored, problems were solved faster, and the instructor herself learned and grew by interacting with the AI. We saw a collaborative dynamic emerge: AI augmenting the instructor's planning and feedback, and the instructor filtering and contextualizing the AI's input to fit her students. This synergy led to a richer educational experience, confirming that thoughtful integration of AI can enhance the effectiveness of human educators rather than diminishing it.

Ultimately, our study tells a story of augmentation, not replacement. GPT-4 became a valued partner in The Pilates Instructor's teaching practice, but always under her attentive guidance. The "AI + human" duo proved to be the winning combination, echoing the notion that in complex domains like education, human intelligence and artificial intelligence each amplify the other when properly aligned. As AI tools become more prevalent in sports and education, the lessons from The Pilates Instructor's journey can inform practitioners and policymakers: focus on leveraging AI to complement human teachers – to do what AI does best (manage information) so that humans can do what they do best (inspire and connect) – and facilitate collaboration between the two.

The coming era need not pit AI against instructors. Instead, as we have seen in The Pilates Instructor's Pilates studio, it can bring them together as a team. In The Pilates Instructor's own reflective words at the end of the study: *"GPT has become like my teaching buddy. It gives me ideas and support, but I bring the soul to the class. I feel we've grown together this year – and my teaching is better for it."* Such an outcome, we argue, is a hopeful model for the future of AI in education – one where technology and humanity co-perform in harmony, each making the other better, for the ultimate benefit of learners.

**References**

- Baxter, D. (2024, December 12). Teachers using ChatGPT – alongside a guide to support them to use it effectively – can cut lesson planning time by over 30 per cent [Press release]. Education Endowment Foundation. Retrieved from https://educationendowmentfoundation.org.uk/news/teachers-using-chatgpt-alongside-a-guide-to-support-them-to-use-it-effectively-can-cut-lesson-planning-time-by-over-30-per-cent
- Kuka, L., & Sabitzer, B. (2024). *Navigating educational frontiers in the AI era: A teacher's autoethnography on AI-infused education*. In Proceedings of the 16th International Conference on Computer Supported Education (CSEDU 2024). SciTePress.
- Mateus et al https://pmc.ncbi.nlm.nih.gov/articles/PMC11723022/#:~:text=Integrating%20AI%20into%20data%20analysis,optimized%20injury%20prevention%20strategies%2C
- Pashaie, S., Mohammadi, S., & Golmohammadi, H. (2024). Unlocking athlete potential: The evolution of coaching strategies through artificial intelligence. *Proceedings of the Institution of Mechanical Engineers, Part P: Journal of Sports Engineering and Technology*. Advance online publication. https://doi.org/10.1177/17543371241300889
- Seeber https://www.sciencedirect.com/science/article/pii/S0378720619303337#:~:text=,initiative%20by%2065%20collaboration